\documentclass{mem}
\usepackage{natbib}\usepackage{txfonts}\usepackage{balance}
\usepackage{graphicx}
\usepackage{txfonts}
\usepackage[a4paper]{hyperref}
\idline{79}{3}
\begin{document}
\def\teff{$T\rm_{eff }$}
\def\kms{$\mathrm {km s}^{-1}$}

\title{Multiple Stellar Populations in Galactic GCs: Observational Evidence
}

   \subtitle{}

\author{
Giampaolo Piotto\inst{1}
          }

  \offprints{G. Piotto}

\institute{
Dipartimento di Astronomia, Universit\`a di Padova, Vicolo
dell'Osservatorio, 3, I-35122 Padova, Italia -- 
\email{giampaolo.piotto@unipd.it}
}

\authorrunning{Piotto}

\titlerunning{Multiple Populations in GCs}

\abstract{
An increasing number of photometric observations of multiple stellar
populations in Galactic globular clusters is seriously challenging the
paradigm of GCs hosting single, simple stellar populations. These
multiple populations manifest themselves in a split of some
evolutionary sequences of the cluster color-magnitude
diagrams. In this paper we will summarize the observational scenario.

}
\maketitle{}

\section{Introduction}

Globular clusters (GC) have generally been though consisting of coeval
and chemically homogeneous stars.  Color-magnitude diagrams (CMD) of
GCs like NGC 6397 (King et al. 1998, Richer et al. 2007) fully support
this paradigm of GCs hosting simple stellar populations.  However,
there is a growing body of observational facts which challenge this
traditional view.  Since the eighties we know that GCs show a peculiar
pattern in their chemical abundances (Gratton el al. 2004 for a
review). While they are generally homogeneous insofar Fe-peak elements
are considered, they often exhibit large anticorrelations between the
abundances of C and N, Na and O, Mg and Al. These anticorrelations are
attributed to the presence at the stellar surfaces of a fraction of
the GC stars of material which have undergone H burning at
temperatures of a few ten millions K (Prantzos et al. 2007). This
pattern is peculiar to GC stars. Field stars only show changes in C
and N abundances expected from typical evolution of low mass stars
(Gratton et al. 2000, Sweigart \& Mengel 1979, Charbonnel 1994). This
abundance pattern is primordial, since it is observed in stars at all
evolutionary phases (Gratton et al. 2001). Finally, the whole stars
are interested (Cohen et al. 2002).

%\begin{figure*}[t!]
% \vspace*{-2.0 cm}
%\resizebox{\hsize}{!}{\includegraphics[width=1.5in]{n6397.eps}}

%\begin{figure}[t!]
% \vspace*{-2.0 cm}
%\includegraphics[width=2.5in]{n6397.eps}
% \vspace*{-1.0 cm}
% \caption{\footnotesize The CMD of NGC 6397 shows very narrow sequences and a well
% defined TO, supporting the idea that, in general, GCs are populated by coeval and
% chemically homoheneous stars.}
%   \label{fig1}
%\end{figure}

In addition, since the sixties we know that the horizontal branches
(HB) of some GCs can be rather peculiar. In some GCs the HB can be
extended to very hot temperatures, implying the loss of most of the
stellar envelope (see compilation by Recio-Blanco et al. 2006). The
distribution of the stars along the HB can be clumpy, with the
presence of one or more gaps (Ferraro et al. 1998, Piotto et
al. 1999).  This problem, usually known as the {\it the second
parameter} problem, still lacks of a comprehensive understanding: many
mechanisms, and many parameters have been proposed to explain the HB
peculiarities, but none apparently is able to explain the entire
observational scenario. It is well possible that a combination of
parameters is responsible for the HB morphology (Fusi Pecci et
al. 1993). Surely, the total cluster mass seems to have a relevant
role (Recio-Blanco et al. 2006).

\begin{figure}[h!]
{\includegraphics[width=2.6in]{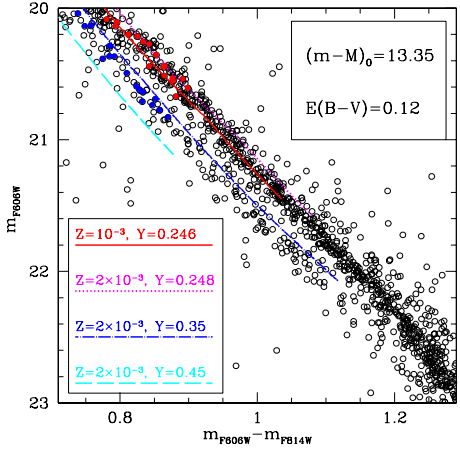}} 
% \vspace*{-1.0 cm}
 \caption{\footnotesize The double MS of $\omega$~Centauri. The bluest MS
 is more metal rich, and it can be reproduced by
 models only assuming a very high He content 0.35$<$Y$<$0.40
 (from Piotto et al. 2005).}
   \label{fig1}
\end{figure}

It is tempting to relate the second parameter problem to the complex
abundance pattern of GCs. Since high Na and low O abundances are
signatures of material processed through hot H-burning, they should be
accompanied by high He-contents (D'Antona \& Caloi 2004). In most
cases, small He excesses up to dY$\sim$0.04 (that is Y$\sim$0.28, assuming the
original He content was the Big Bang one) are expected. While this
should have small impact on colors and magnitudes of stars up to the
tip of the RGB, a large impact is expected on the colors of the HB
stars, since He-rich stars should be less massive. E.g., in the case
of GCs of intermediate metallicity ([Fe/H]~-1.5), the progeny of
He-rich, Na-rich, O-poor RGB stars should reside on the blue part of
the HB, while that of the "normal" He-poor, Na-poor, O-rich stars
should be within the instability strip or redder than it. 
%Actually
%mean HB colors are also influenced by small age differences of 2-3
%Gyr. 
However, within a single GC a correlation is expected between the
distribution of masses (i.e. colors) of the HB-stars and of Na and O
abundances.

In summary, a number of apparently independent observational facts
seems to suggest that, at least in some GCs, there are stars which
have formed from material which must have been processed by a previous
generation of stars.
The questions is: do we have some direct, observational evidence of
the presence of multiple populations in GCs? Very recent discoveries,
made possible by high accuracy photometry on deep HST images, allowed
us to positively answer to this question. In this paper, we will
summarize these new observational facts, and briefly discuss their
link to the complex abundance pattern and to the anomalous HBs.

\section{Direct Observational Evidence of Multiple Populations in GCs}

\begin{figure}[h!]
{\includegraphics[width=2.6in]{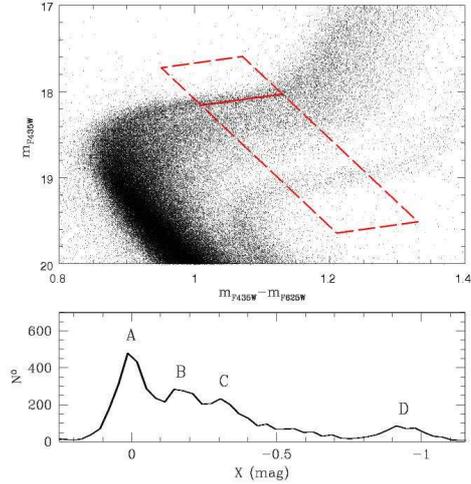}} 
% \vspace*{-1.0 cm}
 \caption{\footnotesize The multiple SGB in Omega Centauri. There are at least 4
 distinct SGBs, plus a small fraction of stars spreaded between SGB-C
 and SGB-D (from Villanova et al. 2007)}
   \label{fig2}
\end{figure}

\begin{figure}[h!]
{\includegraphics[width=2.6in]{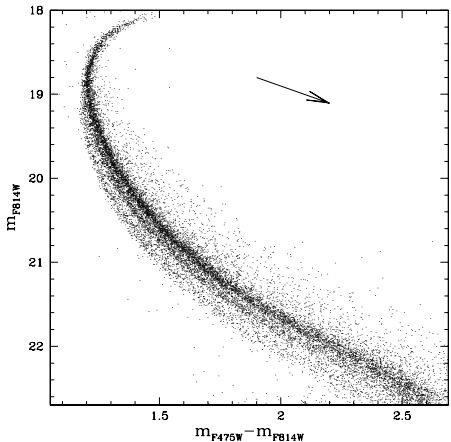}} 
% \vspace*{-1.0 cm}
 \caption{\footnotesize The spectacular triple MS of NGC 2808. Note the narrowness
 of the turnoff. The arrow indicates the reddening line.}
   \label{fig3}
\end{figure}

The first, direct observational evidence of the presence of more than
one stellar population in a GCs was published by Bedin et
al. (2004). Bedin et al. found that, for a few magnitudes below the
turn-off (TO), the main sequence (MS) of $\omega$ Centauri splits in
two (Fig. 1). Indeed, the suspect of a MS split in $\omega$ Cen was
already raised by Jay Anderson in his PhD thesis, but the result was
based on only one external WFPC2 field, and this finding was so
unexpected that he decided to wait for more data and more accurate
photometry to be sure of its reality. Indeed, Bedin et al. (2004)
confirmed the MS split in Jay Anderson field and in an additional ACS
field located 17 arcmin from the cluster center. Now, we know that the
multiple MS is present all over the cluster, though the ratio of blue
to red MS stars diminishes going from the cluster core to its envelope
(Sollima et al. 2007, Bellini et al. 2008, in preparation).

The more shocking discovery on the multiple populations in $\omega$
Cen, however, came from a follow-up spectroscopic analysis that showed
that the blue MS has twice the metal abundance of the dominant red
branch of the MS (Piotto et al.\ 2005).  The only isochrones that
would fit this combination of color and metallicity (cf. Fig.~1) are
extremely enriched in helium ($Y\sim 0.38$) relative to the dominant
old-population component, which presumably has primordial helium.

Indeed, the scenario in $\omega$ Cen is even more complex.  As it is
already evident in the CMD of Bedin et al. (2004), this object has at
least three MSs, which spread into a highly multiple sub-giant branch
(SGB) with at least four distinct components (Fig. 2) characterized by
different metallicities and ages (Sollima et al.\ 2005, Villanova et
al.\ 2007; the latter has a detailed discussion.) A fifth, more
dispersed component is spread between the SGB-C and SGB-D of Fig. 2.

These results reinforced the suspicion that the multiple
MS of $\omega$ Cen could just be an additional peculiarity of an
already anomalous object, which might not even be a GC, but a remnant
of a dwarf galaxy instead.
In order to shed more light on the possible presence of multiple MSs in
Galactic GCs, we undertook an observational campaign with {\sl HST},
properly devised to search multiple sequences at the level of the
upper-MS, turn-off (TO), and SGB. The new data allowed us to show that
the multiple evolutionary sequence phenomenon is not a peculiarity of
$\omega$ Centauri only.

As shown in Fig. 3, also the CMD of NGC 2808 is splitted into three
MSs (Piotto et al. 2007). Because of the negligible dispersion in Fe
peak elements (Carretta et al 2006), Piotto et al. (2007) proposed the
presence of three groups of stars in NGC 2808, with three different He
contents, in order to explain the triple MS of
Fig. 3. These groups may be associated to the three groups with
different Oxygen content discovered by Carretta et al. (2006). These
results are also consistent with the presence of a multiple HB, as
discussed in D'Antona and Coloi (2004) and D'Antona et
al. (2006). Finally, we note that the narrowness of the TO region
displayed by Fig.~3 suggests that the 3 stellar populations of NGC
2808 must have a small age dispersion, much less than 1 Gyr.

\begin{figure*}[t!]
\begin{center}
{\includegraphics[width=5.0in]{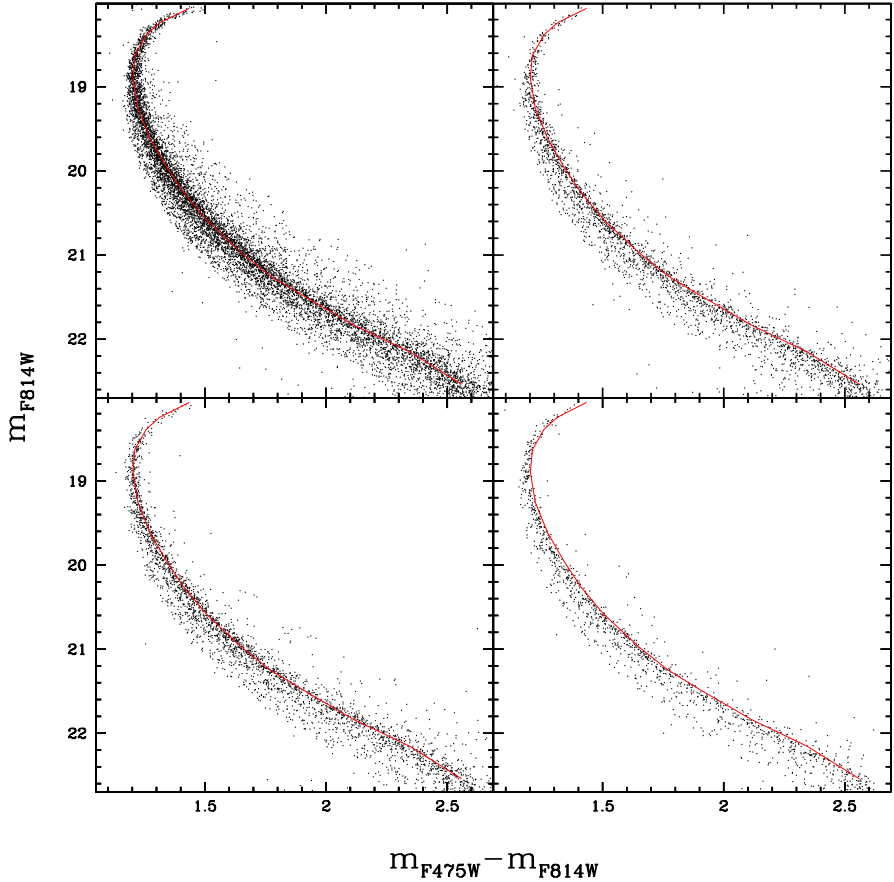}} 
% \vspace*{-1.0 cm}
 \caption{\footnotesize Original, not corrected for differential
 reddening CMD of NGC 2808 in the four quadrants of the ACS field. The
 triple MS is well visible, and with exactly the same shape in all
 quadrants. The small offsets of the fiducial line of the reddest MS
 indicates the level of differential reddening within the ACS field.}
   \label{fig4}
\end{center}
\end{figure*}

At the meeting, someone raised the suspect that the multiple sequences
in NGC 2808 may be an artifact of differential reddening. Indeed, it
is well known since Walker (1999) that NGC 2808 suffers a small amount
of differential reddening.  Bedin et al. (2000) calculated a
differential reddening of the order of 0.02 magnitudes in E(B-V) from
the cluster center to $\sim$400 arcsec from the center. The split of
the sequences in Fig. 3 is more than an order of magnitude larger than
this value. In any case, being aware of the presence of differential
reddening, we corrected for it.  The CMD of NGC 2808 published in Piotto et
al. (2007) and reproduced in Fig.~3 has been corrected for
differential reddening, adopting the procedure described in Sarajedini
et al. (2007), which allows also to correct for small, residual 
spatial variation of the photometric zero point which affects the ACS data
(see Sarajedini et al. 2007 for more details). There are additional
arguments against the differential reddening argument. First of all,
the TO-SGB regions of the CMD are very narrow (this is the region of
the CMD which should be mostly affected by differential reddening, as
shown by the reddening line plotted in Fig. 3). The three MSs of
Fig. 3 tend to diverge from the TO to fainter magnitudes, while
differential reddening would simply cause a shift of the sequences
parallel to the reddening line. Finally, differential
reddening tends to randomly broaden the CMD sequences, not to create
coherent features, which, we note, are exactly the same everywhere
within the ACS field, as shown in Fig. 4.

Also NGC 1851 must have at least two, distinct stellar populations. In
this case the observational evidence comes from the split of the SGB
in the CMD (Fig. 5) of this cluster (Milone et al. 2007). Would the
magnitude difference between the two SGBs be due only to an age
difference, the two star formation episodes should have been
separated by at least 1 Gyr. However, as shown by Cassisi et
al. (2007), the presence in NGC 1851 of two stellar populations, one
with a normal $\alpha$-enhanced chemical composition, and one
characterized by a strong CNONa anticorrelation pattern could
reproduce the observed CMD split. In this case, the age spread between
the two populations could be much smaller, possibly consistent with
the small age spread implied by the narrow TO of NGC 2808. In other
terms, the SGB split would be mainly a consequence of the metallicity
difference, and only negligibly affected by (a small) age
dispersion. Cassisi et al. (2007) hypothesis is supported by the
presence of a group of CN-strong and a group of CN-weak stars
discovered by Hesser et al. 1982, and by a recent work by Yong and
Grundahl (2007) who find a NaO anticorrelation among NGC 1851 giants.

\begin{figure}[h!]
{\includegraphics[width=2.6in]{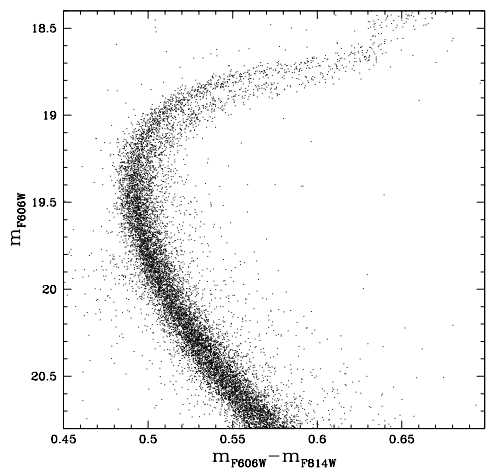}} 
% \vspace*{-1.0 cm}
 \caption{\footnotesize The double SGB in NGC 1851. The two SGBs are separated by
 about 0.12 magnitudes in F606W (from Milone et al. 2007)}
   \label{fig5}
\end{figure}

\begin{figure}[h!]
{\includegraphics[width=2.6in]{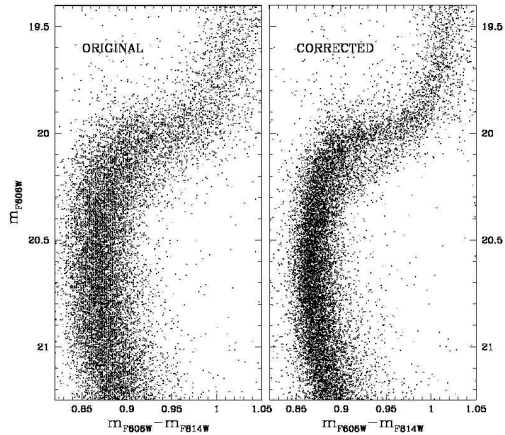}} 
% \vspace*{-1.0 cm}
 \caption{\footnotesize After the correction for differential reddening, also NGC
 6388 shows a double SGB (Piotto et al. 2008, in preparation).}
   \label{fig6}
\end{figure}

NGC 1851 is considered a sort of prototype of bimodal HB clusters.
Milone et al. (2007) note that the fraction of fainter/brighter SGB
stars is remarkably similar to the fraction of bluer/redder HB
stars. Therefore, it is tempting to associate the brighter SGB stars
to the CN-normal, s-process element normal stars and to the red HB,
while the fainter SGB should be populated by CN-strong, s-process
element-enhanced stars which should evolve into the blue HB. In this
scenario, the faint SGB stars should be slightly younger (by a few
$10^7$ to a few $10^8$ years) and should come from processed material
which might also be moderately He enriched, a fact that would help
explaining why they evolve into the blue HB. By studying the cluster
MS, Milone et al. (2007) exclude an He enrichment larger than
$\Delta$Y=0.03), as expected also by the models of Cassisi et
al. (2007). Nevertheless, this small He enrichment, coupled with an
enhanced mass loss, would be sufficient to move stars from the red to
the blue side of the RR Lyrae instability strip.  Direct spectroscopic
measurements of the SGB and HB stars in NGC 1851 are badly needed.

There is at least another cluster which undoubtedly shows a split in
the SGB: NGC 6388 (Piotto et al. 2008, in preparation). Figure 6 shows
that, even after correction for differential reddening, the SGB of NGC
6388 closely resembles the SGB of NGC 1851. NGC 6388, as well as its
twin cluster NGC 6441, are two extremely peculiar objects. Since Rich
et al. (1997), we know that, despite their high metal content, higher
than in 47 Tucanae, they have a bimodal HB, which extends to extremely
hot temperatures (Busso et al. 2007), totally un-expected for this
metal rich cluster. NGC 6388 stars also display a NaO anticorrelation
(Carretta et al. 2007). Unfortunately, available data do not allow us
to study the MS of this cluster, searching for a MS split. Hopefully,
new data coming from the HST program GO11233 should help to constrain
the MS width, and therefore the He dispersion.  In this context, it is
worth noting that Caloi and D'Antona (2007), in order to reproduce the
HB of NGC 6441, propose the presence of three populations, with three
different He contents, one with an extreme He enhancement of
Y=0.40. Such a strong enhancement should be visible in a MS split, as
in the case of $\omega$ Cen and NGC 2808.  A strong He enhancement and
a consequent MS split may also apply to NGC 6388, because of the many
similarities with NGC 6441.

One more cluster, M54, shows a complex CMD (see, e.g., Siegel et al.\
2007).  This cluster has been shown, however, in too many papers to
cite here, to be a part of the Sagittarius dwarf galaxy that is in
process of merging into the Milky Way, and very possibly the 
nucleus of that galaxy.  Actually, it is still matter of debate which
parts of the CMD of M54 represent the cluster population and which
ones are due to the Sagittarius stars.  M54 may have a complex stellar
populations as $\omega$ Cen, though this fact will be much harder to
demonstrate.

%\begin{figure}[h!]
%{\includegraphics[width=2.6in]{mackey.eps}} 
% \vspace*{-1.0 cm}
% \caption{\footnotesize The double main sequence turn off of the Large Magellanic
% Cloud cluster NGC 1846 (from Mackey and Broby Nielsen 2007).}
%   \label{fig6}
%\end{figure}

Finally, we note that the multiple population phenomenon in star
clusters may not be confined only to Galactic GCs. Mackey \& Broby Nielsen 
(2007) suggest the presence of two populations with an age
difference of $\sim$300 Myr in the 2 Gyr old cluster NGC 1846 of the
Large Magellanic Cloud (LMC). In this case, the presence of the two
populations is inferred by the presence of two TOs in the CMD. These
two populations may either be the consequence of a tidal capture of two
clusters or NGC 1846 may be showing something analogous to the multiple
populations identified in the Galactic GCs. NGC 1846 might not be an
exception among LMC clusters. Vallenari et al. (1994) already
suggested the possibility of the presence of two stellar populations
in the LMC cluster NGC 1850. A quick analysis of the CMDs of about 50
clusters from ACS/HST images shows that about 10\% of them might show
evidence of multiple generations (Milone et al. 2008, in preparation).

\section{Discussion}

So far, we have identified four Galactic globular clusters for which we
have a direct evidence of multiple stellar populations, and they are
all quite different:

%\begin{enumerate}

%\item
1) In $\omega$ Centauri ($\sim 4\times10^6$ $M_\odot$), the different populations
manifest themselves both in a MS split (interpreted as a split in He
and metallicity abundances) and in a SGB split (interpreted in terms
of He, metallicity, and age variations $>1$ Gyr), which implies at least
four different stellar groups within the same cluster, which formed in
a time interval greater than 1 Gyr. Omega Centauri has also a 
very extended HB (EHB), which extends far beyond 30.000K.

%\item
2) In NGC 2808 ($\sim1.6\times10^6$ $M_\odot$), the multiple generation
of stars is inferred from the presence of three MSs (also in this
case interpreted in terms of three groups of stars, with different He
content), possibly linked to three stellar groups with different
oxygen abundances, and possibly to the multiple HB. The age difference
between the 3 groups is significantly smaller than 1 Gyr. Also NGC 2808
has an EHB, extended as much as the HB of $\omega$ Cen. It shows an
extended NaO anticorrelation.

%\item
3) In the case of NGC 1851 ($\sim1.0\times10^6$ $M_\odot$), we have
evidence of two stellar groups from the SGB split. It is difficult to
establish the age difference between the two stellar populations
without a detailed chemical abundance analysis. However, available
observational evidence seems to imply that the SGB split may be due to a
difference in CN, Na, O, and s-process elements, while the age
difference could be small (e.g. as small as in the case of NGC
2808). On the other hand, if the
SGB split would be due only to age, the two star formation episodes 
should have happened with a time separation of $\sim 1$ Gyr.
From the analysis of the cluster CMD, there seems to be no MS
split, which would imply a small He spread, if any ($\Delta Y <
0.03$). The cluster has no EHB, but it shows a bimodal HB.
It shows a NaO anticorrelation.

%\item
4) In NGC 6388 ($\sim1.6\times10^6$ $M_\odot$) we have evidence of two
stellar groups from a SGB split. With the available observational
data it is not possible to establish whether there is a split in the
MS of this GC. NGC 6388 has an EHB, possibly as extended as in the cases of
NGC 2808 and $\omega$ Cen.
It shows an extended NaO anticorrelation.

%\end{enumerate}

Another massive ($\sim2.0\times10^6$ $M_\odot$) GC, M54,
is suspected to host multiple populations, though the analysis is
strongly hampered by the contamination of the Sagittarius galaxy. Also
M54 has an EHB, similar to the HB of NGC 2808 and $\omega$ Cen (Rosenberg et
al. 2004).

At least one LMC intermediate age cluster shows a population split at
the level of the TO: NGC 1846. This is a massive clusters, among the
most massive LMC clusters according to Chrysovergis et al. (1999),
though probably not as massive as the above Galactic GCs (a more
accurate mass estimate for this cluster is needed). Other LMC clusters
are suspected to shows a similar TO splitting.

Many GCs are clearly not simple, single-stellar-population
objects. The emerging evidence is that the star-formation history can
vary strongly from GC to GC, and that, GCs are able to produce very
unusual objects, as no such He-rich MS stars have ever been found
elsewhere.  At the moment, we can note that the three GCs in which
multiple generations of stars have been clearly identified (Omega Cen,
NGC 2808, and NGC 1851), and the two other GCs expected to contain
more than one stellar generation (NGC 6388 and NGC 6441: Fig 6, Caloi \&
D'Antona 2007, Busso et al. 2007) are among the ten most massive GCs
in our Galaxy. This evidence suggests that cluster mass might have a
role in the star-formation history of GCs.

Reconstruction of this star-formation history requires a a better
understanding of the chemical enrichment mechanisms, but the site of
hot H-burning remains unclear. There are two requisites: (i)
temperature should be high enough; and (ii) the stars where the
burning occurs should be able to give back the processed material to
the intracluster matter at a velocity low enough that it can be kept
within the GC itself (a few tens of km/s).  Candidates include: (i)
Massive ($M>10M_\odot$) rotating stars (Decressin et al.  2007);
(ii) the most massive among the intermediate mass stars undergoing hot
bottom burning during their AGB phase (Ventura et al. 2001). The two
mechanisms act on different timescales ($10^7$ and $10^8$ yr,
respectively), and both solutions have their pros and cons. The
massive star scenario should avoid mixture of O-poor, Na-rich material
with that rich in heavy elements from SNe, while it is not clear how
the chemically processed material could be retained by the
proto-cluster in spite of the fast winds and SN explosions always
associated to massive stars.  Producing the right pattern of
abundances from massive AGB stars seems to require considerable fine
tuning. In addition, both scenarios require that either the IMF of GCs
was very heavily weighted toward massive stars, or that some GCs
should have lost a major fraction of their original population (Bekki
and Norris 2006), and then may even be the remnants of tidally
disrupted dwarf galaxies, as suggested by the complexity in the CMD of
$\omega$ Cen and M54.

Also in view of these problems, it will be important to understand
the role of the capture of field stars by massive forming globular
clusters, as suggested by Kroupa (1998), see also Fellhauer et al. 2006).

The observational scenario is becoming more complex, but, the new
results might have indicated the right track for a comprehensive
understanding of the formation and early evolution of GCs.  We are
perhaps for the first time close to compose what has been for decades
and still is a broken puzzle, which includes the cluster star chemical
anomalies, the HB peculiarities, and, now, the multiple evolutionary sequences.

\begin{acknowledgements}
I wish to warmly thank J. Anderson, Andrea Bellini, Luigi R. Bedin,
Ivan R. King, Antonino P. Milone, without whom most of the results
presented in this review would not have been possible. I wish also to
thank Sandro Villanova for his help in the spectroscopic investigation
of Omega Centauri main sequence and SGB stars. A special thanks to
Alvio Renzini and Raffaele Gratton for the many enthusiastic
discussions on the subject of multipopulations in globular clusters.

I will for ever be grateful to Vittorio, who, more than anyone else,
gave me the opportunity to appreciate the luck we have to do
research in Astronomy. Ciao Vittorio.
\end{acknowledgements}

\bibliographystyle{aa}

\begin{thebibliography}{}

\bibitem[Bedin(2004)]{B04} Bedin, L. R., Piotto, G., Anderson, J.,
Cassisi, S., King, I. R., Momany, Y., \& Carraro, G. 2004,
ApJ (Letters), 605, L125
\bibitem[Bekki(2006)]{bk06} Bekki, K., \& Norris, J.\ E. 2006,
  ApJ (Letters), 637, L109
\bibitem[Bedin(200)]{b00} Bedin, L.\ R., Piotto, G., Zoccali, M.,
  Stetson, P.\ B., Saviane, I., Cassisi, S., \& Bono, G. 2000,
  A\&A, 363, 159
\bibitem[Busso(2007)]{BU07} Busso et al. (2007), A\&A, 474, 105
\bibitem[Caloi(2007)]{C07} Caloi, V., \& D'Antona, F. 2007, A\&A, 463,
  949
\bibitem[Carretta(2006)]{C06} Carretta, E., Bragaglia, A., Gratton,
  R.\ G., Leone, F., Recio-Blanco, A., Lucatello, S. 2006, A\&A, 450,
  523
\bibitem[Carretta(2007)]{Ca07} Carretta, E. et al. 2007,
  A\&A, 464, 957
\bibitem[Cassisi(2007)]{C07} Cassisi, S., Salaris, M., Pietrinferni,
  A., Piotto, G., Milone, A.\ P., Bedin, L.\ R., Anderson, J. 2007, arXiv0711.3823
\bibitem[Charbonell(1994)]{C94} Charbonnel, C. 1994, A\&A, 282, 811
\bibitem[Chrysovergis(1994)]{C94} Chrysovergis, M., Kontizas, M., \&
  Kontizas, E. 1989, A\&AS, 77, 357 
\bibitem[Cohen(2002)]{C02} Cohen, J.\ G., Briley, M.\ M., Stetson, P.\
  B. 2002, AJ, 123, 2525
\bibitem[Dantona(2004)]{d04} D'Antona, F., \& Caloi, V. 2004, ApJ, 
611, 871
\bibitem[Dantona(2006)]{d06} D'Antona, F., Bellazzini, M., Caloi, V.,
  Pecci, F. Fusi, Galleti, S., Rood, R. T. 2006, ApJ, 631, 868
\bibitem[Decressin(2007)]{d07} Decressin, T., Meynet, G., Charbonnel,
  C., Prantzos, N., \&  Ekström, S. 2007, A\&A, 464, 1029
\bibitem[Fellhauer(2006)]{f06} Fellhauer, M., Kroupa, P., \and Evans,
  N.\ W. 2006, MNRAS, 372, 338
\bibitem[Ferraro(1998)]{f98} Ferraro, F.\ R., Paltrinieri, B., Fusi
  Pecci, F., Rood, R.\ T., Dorman, B. ApJ, 500, 311
\bibitem[Fusi(1993)]{f93} Fusi Pecci, F., Ferraro, F.\ R., Bellazzini,
  M., Djorgovski, S., Piotto, G., Buonanno, R. 1993, AJ, 105, 1145
\bibitem[Gratton(2000)]{G00} Gratton, R., Sneden, C., Carretta, E.,
  Bragaglia, A., A\&A, 354, 169 
\bibitem[Gratton(2001)]{G01} Gratton, R. et al. 2001, A\&A, 369, 87
\bibitem[Gratton(2001)]{G04} Gratton, R.; Sneden, C.; Carretta,
  E. 2004, ARAA, 42, 385
\bibitem[Hesser(1982)]{H82} Hesser, J.~E., Bell, R.\ A., Harris, G.\ L.\
  H., \& Cannon, R.\ D. 1982, AJ, 87, 1470
\bibitem[Kroupa(1998)]{k98} Kroupa, P. 1998, MNRAS, 300, 200
\bibitem[King(198)]{k98} King, I.\ R., Anderson, J., Cool, A., \&
  Piotto, G. 1998, ApJ, 492, L37
\bibitem[Mackey(2007)]{2007} Mackey, A.\ D. \& Broby Nielsen, P. 2007,
  MNRAS, 379, 151 
\bibitem[Milone(2007)]{Mi07} Milone, A. P. et al. 2007, in press, arXiv0709.3762
\bibitem[Piotto(1999)]{P99} Piotto, G., Zoccali, M., King, I.\ R.,
  Djorgovski, S.\ G., Sosin, C., Rich, R.\ M., Meylan, G. 1999, AJ, 118, 1727
\bibitem[Piotto(2005)]{P05} Piotto, G., et al. 2005, ApJ, 621, 777 (P05)
\bibitem[Piotto(2007)]{P07} Piotto, G., et al. 2007, ApJ (Letters), 661, L53 (P07)
\bibitem[Prantzoz(2007)]{Pr07} Prantzos, N., Charbonnel, C., \&
  Iliadis, C. 2007, A\&A, 470, 179
\bibitem[Recio(2006)]{R06} Recio-Blanco, A., Aparicio, A., Piotto, G.,
de Angeli, F., Djorgovski, S. G. 2006, A\&A, 452, 875
\bibitem[Rich(1997)]{R97}   Rich, R.\ M., Sosin, C., Djorgovski, S.\
  G., Piotto, G., King, I.\ R., Renzini, A., Phinney, E. S., Dorman,
  B., Liebert, J., Meylan, G., 1997, ApJ (Letters), 484, L25
\bibitem[Richer(2007)]{r07} Richer, H.\ B. et al. 2007,  arXiv0708.4030
\bibitem[Rosenberg(2004)]{r04} Rosenberg, A., Recio-Blanco, A., \&
  García-Marín, M. 2004, ApJ, 603, 135
%\bibitem[Sandage(1967)]{S67} Sandage, A., \& Wildey, R. 1967, ApJ,
%  150, 469 
\bibitem[Sarajedini(2007)]{ss07} Sarajedini, A. 2007, AJ, 133, 1658
\bibitem[Siegel(2007)]{Si07} Siegel et al. 2007, ApJ (Letters), 667, L57
\bibitem[Sollima(2005)]{S05} Sollima, A., Pancino, E., Ferraro, F.\
        R., Bellazzini, M., Straniero, O., \& Pasquini, L. 2005, ApJ,
        634, 332
\bibitem[Sollima(2005)]{S07} Sollima, A., Ferraro, F.\ R., Bellazzini,
  M., Origlia, L., Straniero, O., Pancino, E. 2007, ApJ, 654, 915
\bibitem[Sweigart(1979)]{S79} Sweigart, A.\ V., \& Mengel, J.\ G. 1979, ApJ, 229, 624
\bibitem[Ventura(2001)]{v01} Ventura, P., D'Antona, F., Mazzitelli, I., 
        \& Gratton, R. 2001, ApJ (Letters), 550, L65
\bibitem[Vallenari(2001)]{v94} Vallenari, A., Aparicio, A., Fagotto,
  F., Chiosi, C., Ortolani, S., \& Meylan, G. 1994, A\&A, 284, 447
\bibitem[Villanova(2007)]{V07} Villanova, S., et al. 2007, ApJ, 663,
  296
%\bibitem[vandenbergh(1967)]{V67} van den Bergh, S. AJ, 72, 70
\bibitem[Walker(1999)]{W99} Walker, A.\ R. 1999, AJ, 118, 432 
\bibitem[Yong(2007)]{Y07} Yong, D. \& Grundahl, F. 2007, arXiv0711.1394


\end{thebibliography}

\end{document}